\documentclass[twocolumn,prb,showpacs,floatfix,superscriptaddress]{revtex4-1}
\usepackage{graphicx}
\usepackage{bm}
\usepackage{amsmath}
\usepackage{amsfonts}
\usepackage{amssymb}
\usepackage[dvipsnames]{xcolor}
\usepackage{float}
\usepackage{cleveref}

\newcommand{\sg}{$Pa \bar{3}$}
\newcommand{\NS}{NiS$_{2}$}
\newcommand{\NSe}{NiSe$_{2}$}
\newcommand{\NSS}{NiS$_{2-x}$Se$_{x}$}

\newcommand{\dSS}{d$_{{\rm S}-{\rm S}}$}
\newcommand{\dSeSe}{d$_{{\rm Se}-{\rm Se}}$}
\newcommand{\dNX}{d$_{{\rm Ni}-X}$}
\newcommand{\dXX}{d$_{X-X}$}
\newcommand{\TIM}{${\rm T}_\mathrm{IM}$}
\newcommand{\cm}{~cm$^{-1}$}

\begin{document}

\title{\textbf{Structural investigation of the insulator-metal transition in \NSS~compounds}}

\author{Garam Han}
\affiliation{Department of Physics and Astronomy, Seoul National University (SNU), Seoul 151-742, Korea}     
\affiliation{Center for Correlated Electron Systems, Institute for Basic Science, Seoul 151-742, Korea}           

\author{Sungkyun Choi}
\email{sc1853@physics.rutgers.edu}
\altaffiliation[Present address: ] {Rutgers Center for Emergent Materials and Department of Physics and Astronomy, Rutgers University, Piscataway, New Jersey 08854, USA}
\affiliation{Department of Physics and Astronomy, Seoul National University (SNU), Seoul 151-742, Korea}     
\affiliation{Center for Correlated Electron Systems, Institute for Basic Science, Seoul 151-742, Korea}           
\affiliation{Max Planck Institute for Solid State Research, Heisenbergstrasse 1, 70569 Stuttgart, Germany}     

\author{Hwanbeom Cho}
\affiliation{Department of Physics and Astronomy, Seoul National University (SNU), Seoul 151-742, Korea}     
\affiliation{Center for Correlated Electron Systems, Institute for Basic Science, Seoul 151-742, Korea}           

\author{Byungmin Sohn}
\affiliation{Department of Physics and Astronomy, Seoul National University (SNU), Seoul 151-742, Korea}     
\affiliation{Center for Correlated Electron Systems, Institute for Basic Science, Seoul 151-742, Korea}           

\author{Je-Geun Park}
\affiliation{Department of Physics and Astronomy, Seoul National University (SNU), Seoul 151-742, Korea}     
\affiliation{Center for Correlated Electron Systems, Institute for Basic Science, Seoul 151-742, Korea}           

\author{Changyoung Kim}
\email{changyoung@snu.ac.kr}
\affiliation{Department of Physics and Astronomy, Seoul National University (SNU), Seoul 151-742, Korea}     
\affiliation{Center for Correlated Electron Systems, Institute for Basic Science, Seoul 151-742, Korea}           

\date{\today}


\begin{abstract}
We report on a combined measurement of high-resolution x-ray diffraction on powder and Raman scattering on single crystalline \NSS~samples that exhibit the insulator-metal transition with Se doping. Via x-rays, an abrupt change in the bond length between Ni and S (Se) ions was observed at the transition temperature, in sharp contrast to the almost constant bond length between chalcogen ions. Raman scattering, a complementary technique with the unique sensitivity to the vibrations of chalcogen bonds, revealed no anomalies in the phonon spectrum, consistent with the x-ray diffraction results. This indicates the important role of the interaction between Ni and S (Se) in the insulator-metal transition. The potential implication of this interpretation is discussed in terms of current theoretical models.
\end{abstract}
\maketitle

\section{Introduction}
\NSS~is an intriguing strongly correlated system showing the transition~\cite{Imada1988} from an insulating \NS~phase to a metallic \NSe~phase via Se doping~\cite{Yao1996,Husmann1996,Abrahams1996,Honig1998,Perucchi,Xu2014}. \NSS~is especially suitable for the study of bandwidth-controlled transitions induced purely by the electron-electron correlation. Specifically, the insulator-to-metal (IM) transition is not obscured by a structural transition, as it preserves its parent crystal symmetry by connecting two isostructural and isoelectronic end compounds. Thus, \NSS~provides a rare opportunity to directly compare experimental and theoretical results.

\begin{figure}[t]
\includegraphics[width=\linewidth]{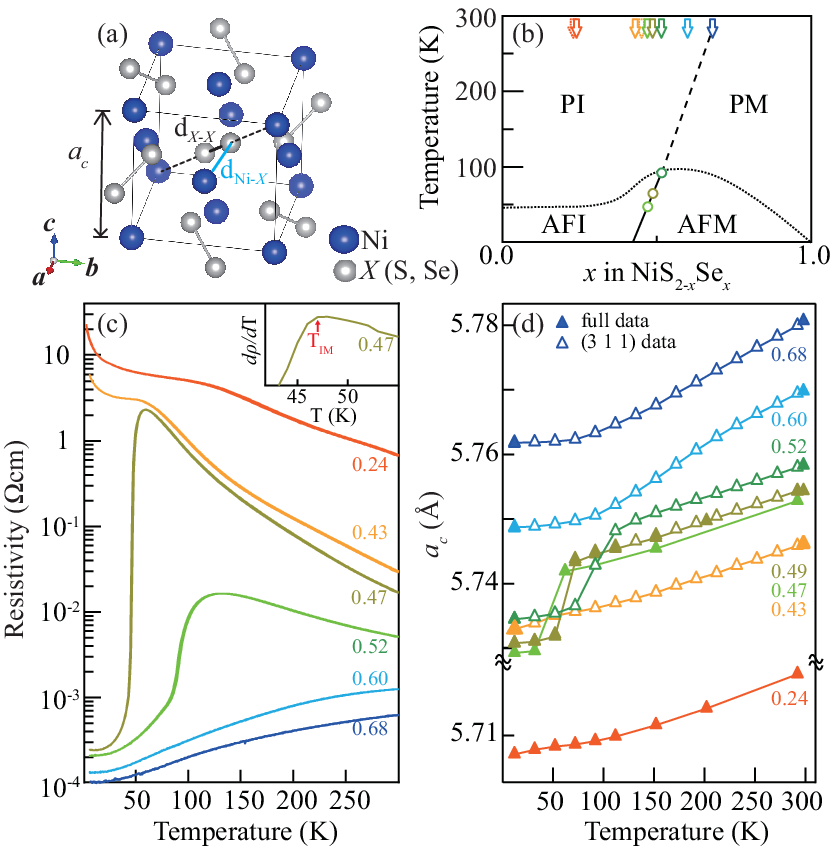}
\caption {(color online) (a) Representative refined pyrite crystal structure of~\NSS~at 292~K ($x$~=~0.49) with a solid blue line representing \dNX~and solid black line representing \dXX. (b) Schematic phase diagram of \NSS~ from $x$~=~0 to 1 with circular symbols obtained from our resistivity data in the \TIM~line (with a solid black line from a linear fit). The transition is of the first-order at the antiferromagnetic boundary (a dotted black line), whereas a crossover-like transition exists in the paramagnetic states (a dashed black line) based on previous neutron works~\cite{Matsuura2000}. AFI (AFM) refers to an antiferromagnetic insulator (metal). The same compositions of compounds used for resistivity measurements in (d) are color-coded in (b) with solid arrows; dotted arrows represent compositions used for Raman measurements, as shown in Fig.~\ref{fig:Raman:TD}. (c) Resistivity data from powder samples. The inset shows the differentiation of resistivity data by temperature, with a clear peak seen at about 47~K, the \TIM. (d) Fitted cubic lattice parameter ($a_{c}$) in the cubic symmetry from powder XRD measurements. Filled (empty) symbols are obtained from fits using the full (focused) 2$\theta$ range of XRD data.}
\label{fig:LP}
\end{figure}

\NSS~has a pyrite structure composed of Ni and chalcogen dimer ions, whose centers of masses are located in the face-centred cubic lattices and at the center of edges of the unit cell, respectively [see Fig.~\ref{fig:LP}(a)]. Its electronic structure can be understood in terms of chalcogen dimer bonding that accepts two electrons from Ni ions, forming the Ni$^{2+}$ state with two electrons occupying the $e_{g}$ orbitals~\cite{Honig1998}. The doping of larger Se ions alters the bond distances as well as the hybridization, stabilizing distinct electronic and magnetic phases [as denoted in the phase diagram in Fig.~\ref{fig:LP}(b)].

The transition in \NSS~can be explained with respect to the $p$-$d$ hybridization in a charge transfer (CT) insulator \NS~\cite{Fujimori1996, Krishnakumar2003}, in which the chalcogen $p$-band is located between upper and lower Hubbard bands of Ni $d$ orbitals. With the substitution of the larger Se ions (than S ions), the $p$-$d$ hybridization becomes stronger and the hopping energy thus increases. This closes the CT gap, stabilizing the metallic phase via the transition~\cite{Matsuura:BD1996, Moon}.

Recently, a new microscopic mechanism~\cite{Kunes2010} involving the $X$ - $X$ ($X$~=~S, Se) dimer in \NSS~has been suggested. In this model, a $p_\sigma^{*}$-band of \NSS~is located below the upper Hubbard band and forms a $p$-band gap with the rest of the $p$-band across the Fermi level. As the length of Se - Se dimer is longer than that of the S - S dimer, the Se substitution for S ions reduces the size of the $p$-band gap and finally leads to the metallic state.

Given this inconsistency in the underlying microscopic mechanism and the controversial role of chalcogen dimers in the transition~\cite{Moon,Kunes2010,Schuster2012,XAS:HJN}, it would be of interest to examine the crystal structure and its dynamics. The literature suggests a strong charge-lattice coupling, such as the sudden volume change observed at the transition temperature~\cite{Miyadai1992,Matsuura2000}. However, to the best of our knowledge, detailed structural investigations of \NSS~with doping and temperature have yet to be made.

Here we present a comprehensive analysis by applying a combined experimental technique of high-resolution x-ray diffraction (XRD) and Raman scattering to \NSS~compounds, to investigate their crystal structures and dynamics. At the transition temperature, XRD measurements showed a sudden contraction of the cubic lattice parameter. A similar trend in the distance between Ni and chalcogen ions upon both Se doping and temperature was also observed. This is in sharp contrast to the gradual change in distance between chalcogen ions. Results from Raman scattering study confirm absence of anomalies in the phonon spectra from the dimers. These consistent results indicate the important role of the interaction between Ni and chalcogen ions in the transition, as opposed to that between chalcogen dimers, more in agreement with the CT mechanism.

The paper is organized as follows. Resistivity data from high-quality powder sample of \NSS~are presented in Sec.~\ref{sec:res}; the results show the evolution of electronic properties over a wide range of doping concentrations. The crystal structures of the same samples were carefully determined with XRD, as presented in Sec.~\ref{sec:pXRD}, followed by Raman scattering analysis of the two single crystals in Sec~\ref{sec:Raman}, in which the change of the Raman-active chalcogen phonons is examined. Their implications of the IM transition are discussed in Sec.~\ref{sec:discussion}. Finally, conclusions are summarized in Sec.~\ref{sec:summary}.

The appendices contain the full details of sample growth in Appendix~\ref{app:sample}, resistivity measurements in Appendix~\ref{app:res}, powder XRD measurements in Appendix~\ref{app:pXRD}, refined crystal structures in Appendix~\ref{app:str}, sample characterizations of single crystals used in Raman measurements in Appendix~\ref{app:Raman:char}, technical details for Raman scattering in Appendix~\ref{app:Raman}, and the effect of beam-heating on Raman measurements in Appendix~\ref{app:Raman:BH}. Tables provide additional information on the refined crystal structures and positions of phonon peaks.

\section{Resistivity}
\label{sec:res}
\begin{figure}[t]
\begin{center}
\includegraphics[width=\linewidth]{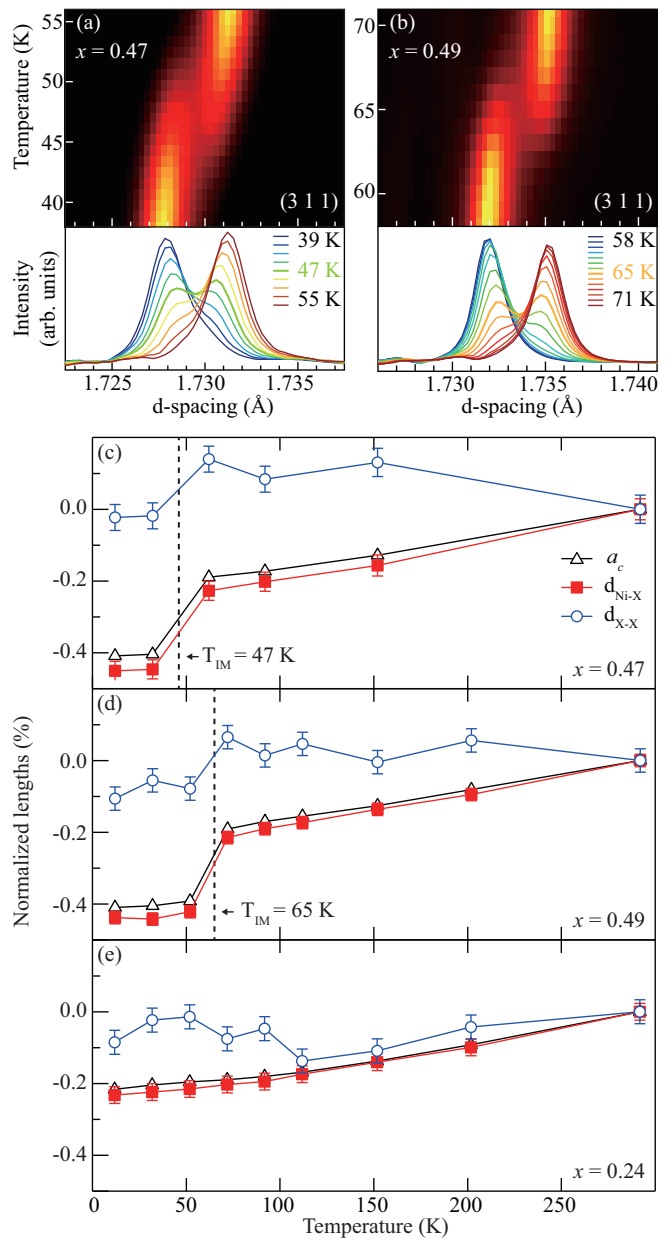}
\end{center} \caption {(color online) (a), (b) Temperature-dependent XRD data for the (3 1 1) peak for $x$~=~0.47 and ~0.49, clearly showing the first-order transition. Cu K$_{\alpha2}$ is subtracted for simpler visualization; small offset angles (-0.04 and -0.11, respectively based on those from the full XRD data) are subtracted in the nominal zero 2$\theta$ angle, before converting to the d-spacing. (c-e) Evolution of $a_{c}$, \dNX~and \dXX~with temperature for $x$~=~0.47,~0.49 and~0.24, respectively.}
\label{fig:XRD_T}
\end{figure}

Figure~\ref{fig:LP}(c) shows the resistivity data from powder samples, categorized into three regimes: insulating ($x$~$\leq$0.43), transition (0.47~$\leq$~$x$~$\leq$~0.52), and metallic ($x$~$\geq~$0.6)~\cite{Matsuura2000}. At lower doping, the samples showed simple insulating behavior with a small hump, suggestive of the effect from a short-range magnetic order~\cite{Matsuura2000}. Resistivity data corresponding to intermediate doping revealed a clear first-order IM transition at $x$~=~0.47 within a narrow temperature range of about 5~K at around 47~K [confirmed as the percolation of two phases in powder XRD, as shown in Figs.~\ref{fig:XRD_T}(a, b)]. This was followed by a crossover-like transition at $x$~=~0.52 (from T = 80~K to 120~K) that became metallic at higher Se doping. The transition at T~=~0~K is expected at $x$~=~0.42, based on extrapolation from a fitted line using three data points from resistivity and XRD data [see the three circular symbols in Fig.~\ref{fig:LP}(b)]. The \NSS~structures of the same powder samples were examined by high-resolution XRD, as explained in the next section.

\section{Powder X-ray diffraction}
\label{sec:pXRD}
High resolution x-ray diffraction (XRD) can provide the structural information in \NSS~associated with the IM transition. Synthesized high-quality powder samples [as described in Appendix~\ref{app:sample}] were examined over a wide range of doping concentrations and temperatures [see solid arrows in Fig.~\ref{fig:LP}(b) for the doping range covered].

\begin{figure}[t]
\begin{center}
\includegraphics[width=\linewidth]{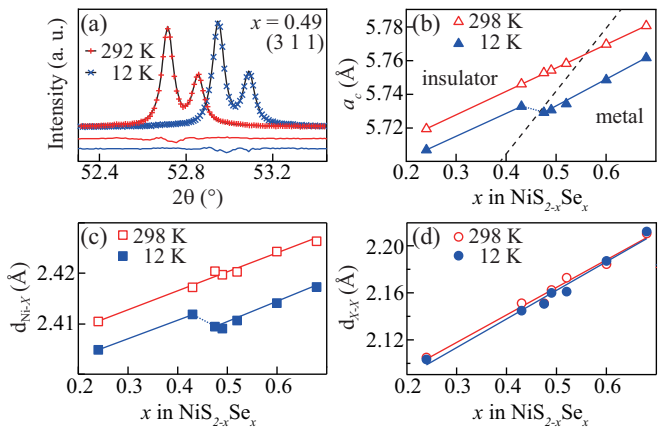}
\end{center} \caption {(color online) Representative (3 1 1) peak data and refined parameters from the full XRD data. (a) Representative XRD data collected at 12~K (cross symbols) and 292~K (plus symbols) with fits (black lines) and the difference between them (horizontal solid lines below the data). (b-d) Extracted structural parameters from the data: (b) $a_{c}$, (c) \dNX, (d) \dXX~with fitted solid lines as guides. The dashed line in (b) is included as a guide for the eyes to separate two phases. Data at 292~K were used for $x$~=~0.24, 0.47 and 0.49.}
\label{XRD_doping}
\end{figure}

To map out the lattice parameters in the phase space of Se doping and temperature, we traced the change in the (3 1 1) Bragg peak [denoted as empty symbols in Fig.~\ref{fig:LP}(d)], taking advantages of its stronger diffraction signal and higher 2$\theta$ resolution compared to other peaks. Figure~\ref{XRD_doping}(a) shows typical double peaks (owing to Cu K$_{\alpha1}$ and K$_{\alpha2}$ wavelengths) for the (3 1 1) peak of the~$x$~=~0.49 sample at room-temperature and 12~K. The fitted full-width-half-maximum of the peak was about 0.056${^\circ}$, which indicated that the powder samples were of a high quality.

Figure~\ref{fig:LP}(d) shows the change of the cubic lattice parameter ${a_c}$ ($c$ for cubic) with temperature; a linear dependence was revealed for insulating powder samples ($x$~=~0.24, 0.43 in \NSS) whereas ${a_c}$ for the metallic powder ($x$~=~0.60, 0.68) exhibited parabolic behavior as the temperature approached 100~K. In contrast, two intermediate dopings ($x$~=~0.47, 0.49) revealed a sudden jump in ${a_c}$ at about 47~K and 65~K, respectively, which coincides with the IM transition temperature (\TIM) in the resistivity data [see Fig.~\ref{fig:LP}(c)], reproducing the results in the literature~\cite{Matsuura2000}.

For the detailed structural analysis of the relationship between the structure and the transition, the full range of 2$\theta$ XRD data were collected for all dopings at room-temperature and 12~K [filled symbols at both temperatures in Fig.~\ref{fig:LP}(d)], and for the three selected dopings over the entire temperature range [filled symbols in Fig.~\ref{fig:LP}(d)]. They are presented in Sec.~\ref{sec:pXRD:DD} and Sec.~\ref{sec:pXRD:TD}, respectively.

\subsection{Doping-dependent powder XRD}
\label{sec:pXRD:DD}
Figure~\ref{XRD_doping}(b) shows an evolution of ${a_c}$ with Se doping at both 298 and 12~K. At 12~K, the lattice parameter is greatly reduced, shifting the (3 1 1) peak to the larger 2$\theta$ [see Fig.~\ref{XRD_doping}(a)]. However, this structural modification does not break the cubic crystallographic symmetry as $X$ ions move along the cubic diagonal direction with the doping. Neither peak splitting nor superlattice peaks caused by a lowered symmetry were found within our experimental resolution; thus, the XRD data supported the parent cubic structure. To examine the structural variations such as bond distances, we performed Rietveld refinement (with the Fullprof package~\cite{fullprof}) using a more realistic Se doping based on Vegard's law~\cite{Chen2002,Otero1998} [see Appendix~\ref{app:str} for the detailed procedure and full results from the refinement].

Figures~\ref{XRD_doping}(b-d) show information on the refined structures, such as the evolution of ${a_c}$, the distance between Ni and S (Se) ions (\dNX), and distance between chalcogen ions (\dXX), respectively. With Se substitution into \NS, the chalcogen ions only move along the diagonal direction in an (x, x, x) fractional coordination of the cubic unit cell, whereas nickel ions remain at the origin. Thus, \dNX~and \dXX~are interconnected and are unambiguously determined by $a_c$ and the fractional coordination-x of $X$ ions inside the unit cell~\cite{Heras}. As illustrated in Fig.~\ref{XRD_doping}(b), ${a_c}$ at 298~K increased with the doping due to the larger ionic size of Se than that of S, agreeing well with Vegard's law and indicative of a well-defined solid solution~\NSS. However, ${a_c}$ at 12~K showed an anomaly for metallic phases ($x$~=~0.49 or 0.52), similar to the evolution of \dNX~at 12~K. This is in sharp contrast to \dXX, in which there were no such anomalies with Se doping at both 298~K and 12~K. Refinement results at 298~K in Figs.~\ref{XRD_doping}(b-d) are compatible with previous XRD measurements on \NSS~thin films~\cite{Otero1998}. To cross-check this structural anomaly, temperature-dependent XRD measurements were conducted on the three selected powder samples by scanning the vertical line (the temperature scan) in the phase diagram of Fig.~\ref{fig:LP}(b); the results are presented in Sec.~\ref{sec:pXRD:TD}.

\subsection{Temperature-dependent powder XRD}
\label{sec:pXRD:TD}
Powder XRD measurements were performed on $x$~=~0.47, 0.49 samples with temperature variation; dramatic first-order transitions were revealed in the resistivity data [see Fig.~\ref{fig:LP}(c)]. Figures~\ref{fig:XRD_T}(a-b) show the evolution of a (3 1 1) peak upon heating for $x$~=~0.47 and 0.49, respectively. In Fig.~\ref{fig:XRD_T}(a), the diffraction peak in the insulating phase at a higher d-spacing ($\sim$~1.7312~\AA) at about T~=~55 K, continued to weaken as the temperature was lowered, whereas the same (3 1 1) peak from the metallic phase began to emerge at a smaller d-spacing ($\sim$~1.7279~\AA) at around 47~K. These features coexisted over a narrow temperature range before the former peak at the higher d-spacing disappeared at about T~=~38~K within the experimental error. The observed transition temperature is consistent with \TIM~of $x$~=~0.47~found in the resistivity data [see a sudden drop in resistivity in Fig.~\ref{fig:LP}{c)]. This behavior is qualitatively the same for $x$~=~0.49 as shown in Fig.~\ref{fig:XRD_T}(b), with a higher transition temperature at about 65~K, and is reminiscent of the coexistence of metallic and insulating phases previously studied in VO$_{2}$~\cite{Qazilbash2007} and chromium-doped V$_{2}$O$_{3}$~\cite{Lupi2010}.

As shown in Figs.~\ref{fig:XRD_T}(c-d), \dNX~decreased with the temperature, with a sharp contraction seen near the transition temperature; the decrease in \dXX~is less significant. Thus, \dXX~was relatively more rigid with temperature. The absence of any strong anomaly in \dXX~with temperature is also consistent with the doping-dependent XRD results shown in Fig.~\ref{XRD_doping}(d). Closer examination of the results shown in Figs.~\ref{XRD_doping}(b-c) and Figs.~\ref{fig:XRD_T}(c-d) revealed a \dNX~anomaly in terms of both temperature and doping dependence.

Measurements were repeated with the non-IM sample ($x$~=~0.24) for a comparative analysis; there was no evidence of any anomaly in the parameters examined ($a_{c}$, \dNX~and~\dXX), as summarized in Fig.~\ref{fig:XRD_T}(e). This observation can be considered with respect to the cubic crystal structure with an arrangement of Ni - $X$ - $X$ - Ni ions along the cubic diagonal direction [depicted as a black dashed line in Fig.~\ref{fig:LP}(a)]. The reduction in the cubic volume with Se doping corresponds to direct contraction of the diagonal distance containing \dXX, expecting a subsequent sudden contraction of \dXX. However, our XRD results showed that the lattice contraction accompanies with \dNX~via the transition [the blue solid bond in Fig.~\ref{fig:LP}(a) distinct from the diagonal direction]. This counterintuitive result indicates that the hybridization of Ni and $X$ ions via the transition plays a more significant role than that of the chalcogen dimers.

\begin{figure*}[t]
\begin{center}
\includegraphics[width=\linewidth]{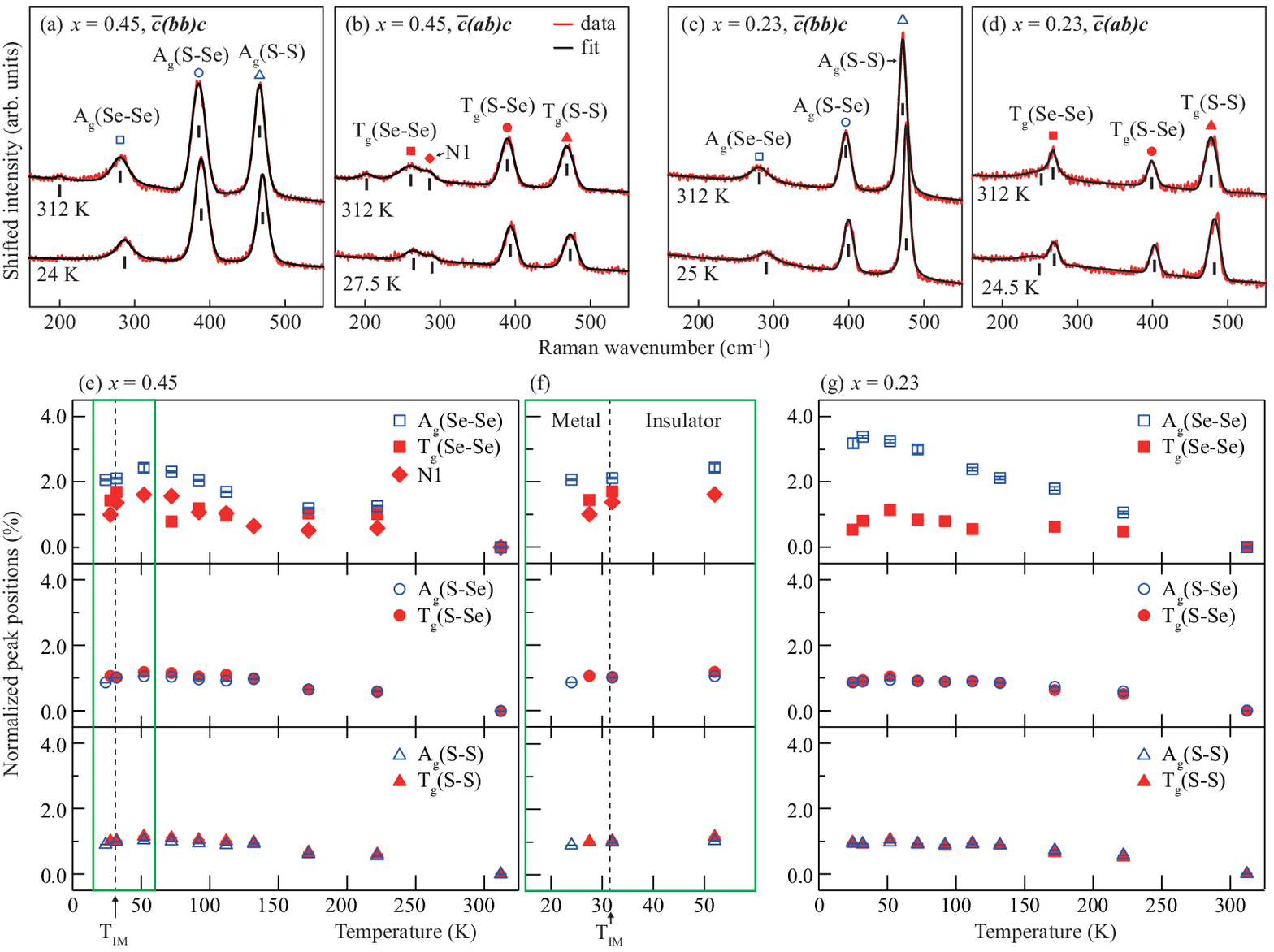}
\end{center} \caption {(color online) (a-d) Raman spectra for the parallel and perpendicular polarization of $x$~=~0.45 and 0.23. (e, g) Normalized fitted peak positions with temperature in $x$~=~0.45 and 0.23, respectively. (f) Zoom view of (e) marked by a green box, for closer examination of the peak shifts near the transition temperature. Symbols for the main peaks in (a-d) match those in (e-g). The fitting error bars shown in (e-g) are very small. The instrumental resolution (1.8\cm) was smaller than the width of vertical bars noted for peak positions in (a-d).}
\label{fig:Raman:TD}
\end{figure*}

\section{Raman scattering on single crystals}
\label{sec:Raman}
To further explore the evolution of the chalcogen dimer through the IM transition, we performed Raman scattering measurements on $x$~=~0.45~(\TIM~$\sim$ 31.5~K) and $x$~=~0.23~(no transition) single crystals [see Appendix~\ref{app:Raman:char} for characterizations of single crystals]. Because Raman scattering is insensitive to the vibrations of nickel ions located at the inversion symmetry position, it is a suitable tool to investigate the dynamics of only chalcogen ions through examining the effect of Se substitution in \NSS. In this regard, it would be intriguing to prove if the changes in electric properties in \NSS~are accompanied by corresponding anomalies in the Raman-active phonon vibrations in the dimer model. Also, Raman measurement may provide evidence of a structural anomaly related to the dimer in \NSS~with doping and temperature, by exploiting its high sensitivity to structural and magnetic transitions~\cite{Postorino2002, Taheri2016}.

However, in the literature, unpolarized Raman measurements of \NSS~\cite{Heras,Lemos1980,Marini:Raman2011,Marini2014} are only reported, mostly at room-temperature. Low-temperature measurements on \NSS~provide deeper insight into the dynamics of chalcogen bonds upon doping and temperature variation via electronic transitions. In particular, polarized analysis with single crystals allows unambiguous assignment of in-phase and out-of-phase stretching vibrations of all possible chalcogen pairs (S - S, S - Se and Se - Se) based on the symmetry argument.

Therefore, polarized Raman measurements were conducted at room-temperature and lower temperatures on two crystals. Representative Raman responses are shown in Figs.~\ref{fig:Raman:TD}(a-d) for two polarizations and two end temperatures [see Figs.~\ref{app:Raman:TD}(a-d) in Appendix~\ref{app:Raman} for the entire temperature data set]. Data were collected from -50 to 1000\cm; however, only the results from 160 to 550\cm~are shown as there were no noticeably strong peaks above 550\cm, apart from weak multi-phonon features.

Group analysis predicts five Raman-active phonons in \NS, A$_g$~+~E$_g$~+~3T$_g$~\cite{Suzuki}. Previous unpolarized Raman measurements on \NS~\cite{Heras,Lemos1980,Marini:Raman2011,Marini2014} revealed A$_g$ (in-phase) and T$_g$ (out-of-phase) stretching modes of the S - S bond at around 470\cm~and two very weak rotational modes (of the dimer dumbbell) E$_g$ and T$_g$ near 270\cm. Once Se ions are substituted, a new pair of phonons from the S - Se bond emerged in the middle energy range [see A$_g$ (S - Se) and T$_g$ (S - Se) in Figs.~\ref{fig:Raman:TD}(a-d)]. A pair of stretching phonons (A$_g$, T$_g$) from Se - Se degenerated inadvertently~\cite{Heras} with weak rotational modes (T$_{g}$, E$_{g}$) from the S - S bond around 270\cm. However, strong A$_g$~(in-phase) and T$_g$~(out-of-phase) stretching modes of all three chalcogen bonds in \NSS~compounds were detectable separately via the polarization measurements, as shown in Figs.~\ref{fig:Raman:TD}(a-d); their phonon energies decreased in the order of S - S, S - Se and Se - Se, as the mass of the chalcogen bond became heavier~\cite{Lemos1980,Heras}.

Raman measurements were performed at various temperatures from 24~K (the base temperature) to 312~K (close to room-temperature); two representative end temperature datapoints are shown in Figs.~\ref{fig:Raman:TD}(a-d). They were obtained using parallel $\bar{\bm c}(\bm b \bm b){\bm c}$ and perpendicular $\bar{\bm c}(\bm a \bm b){\bm c}$ polarizations~\cite{Suzuki}, where the first (second) axis inside the parenthesis refers to the polarisation of incoming (outgoing) light, and the left (right) axis outside the parenthesis corresponds to the incoming (outgoing) direction of light [see Appendix~\ref{app:Raman} for details, including the Raman intensity tensors in Eq.~(\ref{eq:Raman})]. For more accurate determination of peak positions, both inelastic and elastic signals were measured simultaneously, and Raman peaks were calibrated based on the corresponding elastic peak [see Appendix~\ref{app:Raman} for details].

With the room-temperature Raman data (qualitatively consistent with the literature~\cite{Lemos1980,Marini:Raman2011,Marini2014}), the assignment of Raman-active phonons was based on their symmetry relations with respect to the light polarizations, the doping dependence of relative intensities, and the energy shift due to the change in the reduced masses of the dimers~\cite{Lemos1980,Heras,Marini:Raman2011,Marini2014} in two crystals ($x$~=~0.45 and 0.23): A$_g$ and E$_g$ in $\bar{\bm c}(\bm b \bm b){\bm c}$ and T$_g$ phonons in $\bar{\bm c}(\bm a \bm b){\bm c}$.} Compared to $x$~=~0.23, phonons in $x$~=~0.45 showed decreased wavenumbers; this observation can be understood in terms of the increased distance between chalcogen ions. Also, the intensity of the S - Se relative to that of the S - S phonon was larger in the more highly doped crystal, owing to larger population of S - Se bonds [compare Fig.~\ref{fig:Raman:TD}(a) with Fig.~\ref{fig:Raman:TD}(c), and Fig.~\ref{fig:Raman:TD}(b) with Fig.~\ref{fig:Raman:TD}(d)].

To quantitatively determine the positions of phonons, data were fitted with pseudo-Voigt profiles [see Appendix~\ref{app:Raman} for test results with profile functions]; peak positions were normalized by room-temperature values by comparing trends in the normalized peak positions with temperature, as illustrated in Figs.~\ref{fig:Raman:TD}(e-g). There was no evidence of any anomalous feature in phonon wavenumber through the IM transition for $x$~=~0.45~[Figs.~\ref{fig:Raman:TD}(a, b) and (e, f)] and $x$~=~0.23 [see Figs.~\ref{fig:Raman:TD}(c, d, g)] samples; this Raman evidence is consistent with powder XRD results shown in Fig.~\ref{fig:XRD_T}(c-e). The peaks of all three chalcogen bonds shifted monotonically toward a higher energy; this can be understood as a simple hardening of phonons owing to more contracted bonds at a lower temperature. We should mention that the temperature shown here is rather representative due to beam-heating [estimated and explained in Appendix~\ref{app:Raman:BH}], implying that no anomaly is observed {\textit{around}} \TIM~in $x$~=~0.45, similar with the results for $x$~=~0.23. We also observed a new Raman peak denoted as N1 in the $\bar{\bm c}(\bm a \bm b){\bm c}$~polarization for $x$~=~0.45; its temperature dependence was similar with those of other peaks, and thus was not associated with the transition. However, this could indicate an inhomogeneity of Se - Se bonds in the $x$~=~0.45 sample.

It is also worth noting that saturated and softened phonon peaks (most prominent for A$_g$ modes) at lower temperatures in both crystals may be indicative of coupling with an antiferromagnetic magnetic structure [see Fig.~\ref{fig:LP}(b)], referring to a similar Raman feature observed in the magnetoelastic distortion of EuCrO$_{3}$~\cite{Taheri2016}. The peak determination for some data [i.e., 132~K for $x$~=~0.45] is ambiguously affected by a strong ice peak nearby; this increases the uncertainty in the extracted peak positions, and for this reason such peak positions are not displayed in Figs.~\ref{fig:Raman:TD}(e-g). There were no anomalies in the peak widths (not shown) through the transition temperature in $x$~=~0.45 and $x$~=~0.23 samples. For completeness, fitted peak positions are listed in Table~\ref{table:Raman}, and the entire set of Raman data is shown in Fig.~\ref{app:Raman:TD} in Appendix~\ref{app:Raman}.

\section{Discussion}
\label{sec:discussion}
The key observation in this paper is an anomalous contraction of \dNX, in sharp contrast to \dXX~via the IM transition. \dNX~exhibited a noticeable contraction through the transition with respect to both doping concentration and temperature, whereas \dXX~did not show any particular anomaly, as confirmed by both XRD and Raman measurements [Figs.~\ref{XRD_doping}(c-d) and~\ref{fig:XRD_T}(c-e), respectively]. If there is a significant contraction of $X$ - $X$ at the transition temperature, the ion-selective Raman scattering should reveal a sudden change in $X$ - $X$ phonons, such as that observed at 230~K in (chromium-doped) V$_{2}$O$_{3}$~\cite{Tatsuyama:VORaman:1980}. However, our Raman data indicated only a simple hardening of stretching frequencies of chalcogen bonds near the transition temperature, consistent with the XRD data. Therefore, we conclude that there is only a sharp contraction in \dNX~at the transition, with a monotonic change in \dXX~remaining.

In general, the reduction of the interatomic distance tends to increase the bandwidth~\cite{Imada1988}; hence, the decrease in \dNX~increases the bandwidth~\cite{Moon} and/or the $d$-$p$ hybridization~\cite{Fujimori1996}. Given that the level of the \dNX~contraction at the transition is comparable with that of the lattice parameter contraction [such that the volume $a_c^{3}$ changes as shown in Figs.~\ref{fig:XRD_T}(c-d)], the \dNX~contraction appears to be related more strongly to the driving force for the transition than the dimer. If the dimer is a direct driving factor for the IM transition, \dXX~would exhibit the anomaly across the first-order transition temperature; however, our XRD and Raman measurements did not show evidence of this.

The sudden drop in ${a_c}$ (with the volume decrease) with decreasing temperature, which concurs with the transition~\cite{McWhan1970,Jardim2015}, can be understood by the Clausius-Clapeyron relation~\cite{Matsuura2000,Jardim2015}, an equation describing the entropy change between insulating and metallic phases. The calculated entropy change (from the x~=~0.49 data shown in Fig.~\ref{fig:XRD_T}(d)) is compatible with the value from heat capacity measurements~\cite{Sudo1986}, demonstrating the first-order nature of the transition in~\NSS. As this first-order transition can be attributed to various origins, it is important to point out that the entropy change~\cite{Sudo1986} by the volume contraction is tiny [i.e., only about 0.2~\% for the temperature shown in Figs.~\ref{fig:XRD_T}(c-d)], and is comparable with other materials proposed in terms of the electronic-driven transition~\cite{Jardim2015}. Also, our Raman measurements revealed only a slight monotonic shift in the stretching frequencies near the transition temperature, thus decreasing the likelihood of a significant contribution to the entropy (in sharp contrast to (chromium-doped) V$_{2}$O$_{3}$~\cite{Tatsuyama:VORaman:1980}). Therefore, our data favor an electronic origin, as opposed to a dimer-driven mechanism. Furthermore, the CT picture is consistent with a recent observation of a bandwidth-controlled transition by angle-resolved photoemission spectroscopy measurements~\cite{Xu2014,GH:arpes}, explained with respect to a first-order transition at the microscopic scale~\cite{GH:arpes}.

On the other hand, it would be interesting to note a possibly distinct mechanism of a pressure-induced IM transition. Pressure-dependent single crystal XRD~\cite{Fujii1987} and Raman spectroscopy~\cite{Marini:Raman2011} made at room temperature reported potential anomalies of the bond lengths and vibrations of the chalcogen dimers respectively in the context of the transition. This could mean a non-negligible role of the chalcogen dimer in the transition under pressure, which may indicate a different microscopic origin of the pressure-induced transition from those of the doping- and temperature-induced transitions. This counterintuitive complexity further makes these compounds highly intriguing to investigate the detailed microscopic mechanisms of the IM transitions driven by various perturbations.

Although we did not observe any structural transition in the \NSS~doping concentrations studied, it would be worthwhile to investigate the possibility of a structural transition in \NSS. Given that \dSS~in \NS~\cite{ANDRESEN_LP} is very different from \dSeSe~in \NSe~\cite{ANDRESEN_LP}, any periodic modulation induced by substituted Se ions in \NSS~will naturally break the parent cubic symmetry, inducing peak splitting and/or the emergence of new nuclear Bragg peaks. However, our XRD results did not show any evidence of these features within experimental error, despite the high resolution of the measurements. Instead, our data fit well with the parent cubic crystal structure (No.~205,~\sg) having a single Wyckoff site ($8c$) for both S and Se ions [as given in Table~\ref{table:RT}]; this indicates that \NSS~is a well-mixed solid solution, also justifying Vegard's law, i.e., there is a well-defined linear change in the cubic lattice parameter with Se doping in \NSS.

Nevertheless, it might be possible that very weak structural distortions take place beyond our experimental sensitivity. For instance, the d-spacing resolution of our x-ray measurements (to be able to detect the change in the lattice parameter) is of the order of 0.01~\AA, which is still one order bigger than the 0.001~\AA~order from thermal expansion measurements~\cite{Nagata:thermal:LP1976}; thus, the latter technique may be able to resolve very small structural distortions in \NSS, similar with the proposed structural distortion in \NS~at low temperatures~\cite{Nagata:thermal:LP1976}.

Our combined experimental results are consistent and well explain the first order transition; our findings also emphasize that refined crystal structures with well-fit high-quality XRD data can serve as reliable experimental constraints when examining the microscopic mechanisms of the IM transitions, such as that involving \textit{ab initio} band structure calculations.

\section{Summary}
\label{sec:summary}
We performed high-resolution XRD measurements on high-quality polycrystalline samples and Raman scattering experiments on high-quality \NSS~single crystals. A detailed structural study with powder XRD measurements indicated that first-order IM transition accommodated a sudden contraction of \dNX~with a hardly noticeable anomaly in \dXX~, with both doping and temperature variations. The potential impact of chalcogen dimers on the electronic and structural transition were investigated by Raman scattering; only a monotonic trend of hardening of lattice dynamics, without any anomalous signature from the \dXX~dimer phonons, was revealed in a good agreement with XRD results. These observations imply the significant effect of the interaction between Ni and chalcogen ions in the IM transition, as opposed to those involving the dimer, thus suggesting the important role of $p$-$d$ hybridization in the metallization process of \NSS~compounds at low temperatures.

\section{Acknowledgements}
We thank D. J. Song, S. R. Park, G. L. Pascut and B. Keimer for reading the manuscript and providing helpful comments. We also appreciate the technical support from H. Y. Choi and Y. J. Choi at Yonsei University for Physical Property Measurement System measurements, K. Son and G. Sch\"{u}tz for Magnetic Property Measurement System experiments and A. Schulz for Raman measurements at Max Planck Institute for Solid State Research, J. Y. Kwon for four-circle x-ray diffraction measurements and H. Sim for single crystal x-ray diffraction experiments at Center for Correlated Electron Systems (CCES), Institute for Basic Science (IBS), and Y. H. Jung and D. Ahn for synchrotron x-ray diffraction experiments at 9B beamline at Pohang Accelerator Laboratory. The work at the IBS-CCES was funded by Institute for Basic Science in Korea (Grant No. IBS-R009-G1 and IBS-R009-G2). S. C. was also supported by the international postdoctoral scholarship at Max Planck Institute for Solid State Research in Germany.

\appendix

\section{Sample growth}
\label{app:sample}

\begin{figure}
\begin{center}
\includegraphics[width=\linewidth]{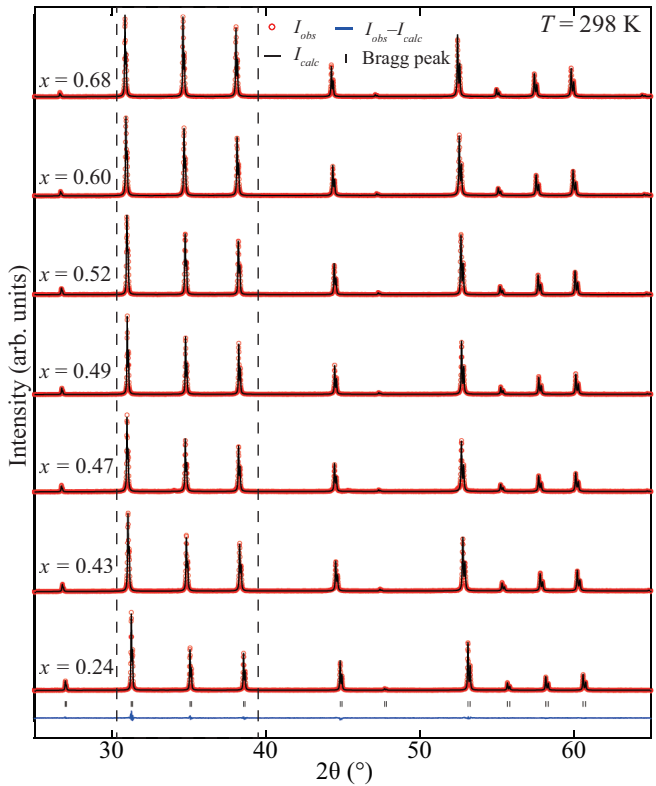}
\end{center} \caption {(color online) Series of XRD patterns on \NSS~($x$~=~0.24, 0.43, 0.47, 0.49, 0.52, 0.60, 0.68) collected at 298~K. Observed XRD data (red circles) and calculated diffraction patterns (black lines) are shown. The difference (a blue solid line) between observed and calculated patterns for $x$~=~0.24 is displayed as a representative data, showing a very good fit. The refinement of all other data was performed in the same manner, with similar fitting results. Note that the data are normalized by an intensity of (2 0 0) peak (near 2$\theta$~$\sim$~31$^{\circ}$) for a better comparison of data collected from different doping concentrations.} \label{app:XRD:RT}
\end{figure}

\begin{figure}
\begin{center}
\includegraphics[width=\linewidth]{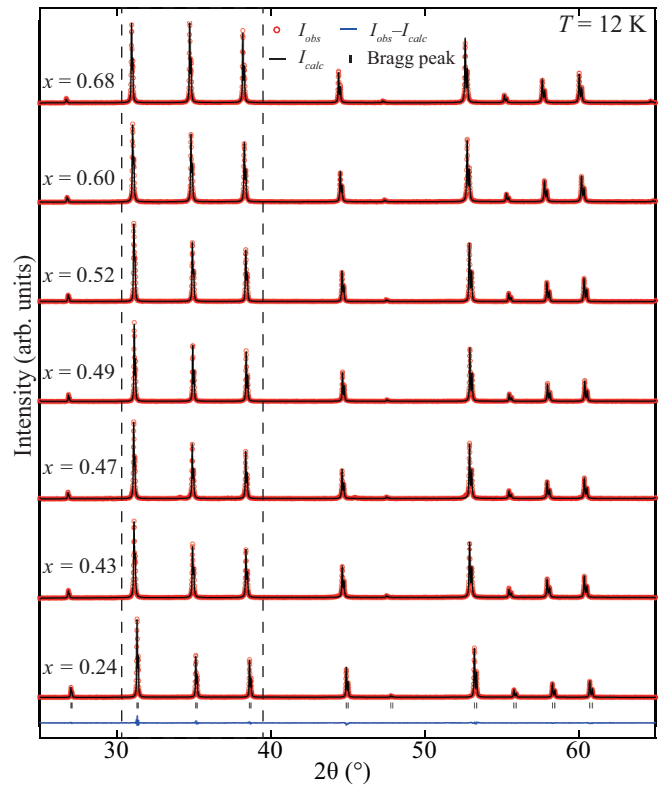}
\end{center} \caption {(color online) Series of XRD patterns on \NSS~($x$~=~0.24, 0.43, 0.47, 0.49, 0.52, 0.60, 0.68) collected at 12~K. Note that the data are normalized by the intensity of (2 0 0) peak for a better comparison.}
\label{app:XRD:LT}
\end{figure}

Series of polycrystalline \NSS~samples ($x$~=~0.24, 0.43, 0.47, 0.49, 0.52, 0.60 and 0.68) were synthesized using a conventional solid-state reaction covering the insulating to metallic phase [see solid arrows in Fig.~\ref{fig:LP}(b)]. Two~grams of starting powder, consisting of Ni (99.99~\%), S (99.9995~\%) and Se (99.999~\%), were mixed well with 3~\% of excess S and Se to compensate an unavoidable evaporation of chalcogen ions during the heating, targeting fully occupied anion ions in powder samples. The pelletized powder was placed in an evacuated silica tube and heated to 720~$^{\circ}$C, at a rate of 20~$^{\circ}$C per hour, inside a box furnace and maintained for 3~days. The resultant powder sample was roughly checked by x-ray measurements to confirm an absence of any secondary phase, followed by a second sintering to improve powder quality. The sintered powder showed very good sample quality without a secondary phase in x-ray measurements.

Single crystals were also grown via a chemical vapour transport method using Cl$_{2}$ gas~\cite{Matsuura2000}. The well-ground starting polycrystalline sample (0.8 g) was sealed under 0.5 atm of Cl$_{2}$ gas in an evacuated silica ampoule (inner diameter: 15~mm; length: 13~$\sim$~14~cm) inside a two-zone furnace. The middle temperature of the ampoule was maintained at about 745~$^{\circ}$C with a temperature gradient of 2~$^{\circ}$C/cm. Single crystals, typically up to 1~$\sim$~2~mm, were synthesized on the tube edge after about a month. The crystals typically exhibited shiny and flat surfaces of (0 0 1) facets\cite{Bouchard1968}, characterized by the Laue diffraction and four-circle XRD measurements (not shown).

\begin{figure}
\begin{center}
\includegraphics[width=\linewidth]{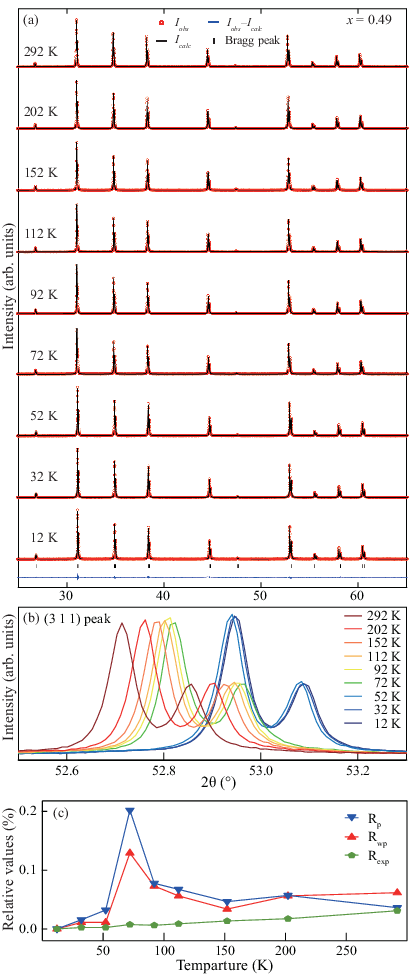}
\end{center}
\caption {(color online) (a) Temperature-dependent XRD data of the $x$~=~0.49~powder sample with refinement results. The difference (a blue solid line) between the diffraction data pattern and calculation at 12~K is shown to represent a good fit. (b) Evolution of the (3 1 1) Bragg peak with temperatures. (c) R${\rm_{p}}$, R${\rm_{wp}}$ and R${\rm_{exp}}$ with relative scales to show how refinements capture a structural anomaly around the first-order IM transition at about 65~K.}
\label{app:XRD:0p49}
\end{figure}

\begin{figure}
\begin{center}
\includegraphics[width=\linewidth]{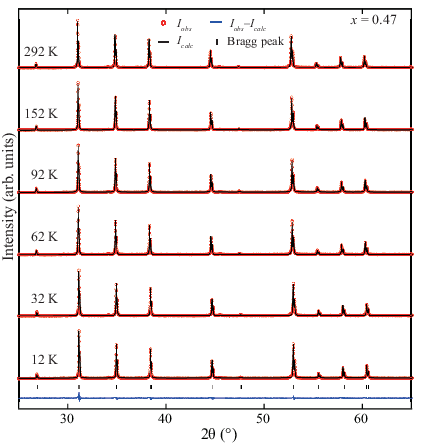}
\end{center}
\caption {(color online) Temperature-dependent XRD data of the $x$~=~0.47~powder sample with refinement results.}
\label{app:XRD:0p47}
\end{figure}

\begin{figure}
\begin{center}
\includegraphics[width=\linewidth]{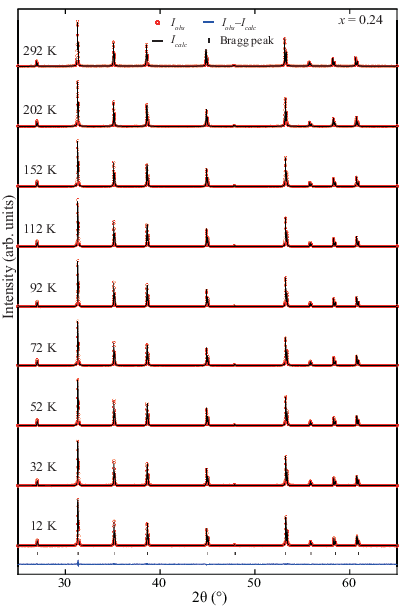}
\end{center}
\caption {(color online) Temperature-dependent XRD data of the $x$~=~0.24~powder sample with refinement results.}
\label{app:XRD:0p24}
\end{figure}

\section{Resistivity}
\label{app:res}
Resistivity measurements of pelletized and sintered powder samples (1 $\times$ 2 $\times$ 0.8~mm$^3$) were performed with Physical Property Measurement System (from Quantum Design). The temperature was scanned at a constant rate of 3~K/min from T~=~5~K to 300~K using a standard four-probe method with gold wires as contacts.

\section{Powder X-ray diffraction}
\label{app:pXRD}
Using high-quality powder samples grown by the method detailed in Appendix~\ref{app:sample}, XRD measurements were performed with a Rigaku D8 advance high-resolution x-ray diffractometer using a Bragg-Brentano geometry with a Cu K$_{\alpha1,2}$ source. No monochromator to screen Cu K$_{\alpha2}$ was used (so Fullprof codewords were applied accordingly), but the Ni filter on the detector ``arm" was used to suppress Cu K$_\beta$ x-ray beams. The following instrumental conditions were applied: 280 mm (radius) diffraction arm; 0.26$^{\circ}$ (0.6 mm) divergence slit; 1.32$^{\circ}$ (3.0 mm) anti-scatter slit; primary Soller slits with 2.5$^{\circ}$ opening. A small mass (approximately 120~mg) of powder sample was well-spread on a circular sample holder to ensure good-quality diffraction patterns. Powder XRD data was collected in a continuous mode at a rate of 1.335~seconds~/~step and a step increment of 0.01$^{\circ}$.

A cryostat was installed for low-temperature measurements. After vacuum pumping of the chamber for several hours, the chamber was cooled to the starting base-temperature (12~K) before beginning the measurements. Notably, the cooling data was qualitatively the same as the heating data, with a marginally lowered transition temperature by about 0.5~K; this difference is consistent with resistivity measurements (not shown). The x-ray diffractometer was calibrated by aligning components before collecting the main data; this improved the 2$\theta$-resolution, particularly in the region of higher 2$\theta$. An aging effect was observed in powder samples, such that magnetic susceptibility at low-temperature increased significantly after several months (not shown) in accordance with observations from a previous study~\cite{NSSpowder:aging}. Thus, XRD measurements were performed soon after synthesizing powder samples. No obvious evidence of degradation of single crystals was observed (over the course of several years).

\section{Structure determination}
\label{app:str}
The full XRD data scan ranged from 20$^{\circ}$ $\leq$ 2$\theta$ $\leq$ 120$^{\circ}$; however, the pattern beyond 2$\theta$ $>$ 65$^{\circ}$ was too weak to be used for the refinements. Thus, for simplicity, diffraction data (red circles) with fitted patterns (black solid lines) are shown only for the range 25$^{\circ}$ $\leq$ 2$\theta$ $\leq$ 65$^{\circ}$ in Fig.~\ref{app:XRD:RT} to Fig.~\ref{app:XRD:0p24}.

In the refinement, a zero-shift and lattice parameters were first found using \textit{Le Bail} fits, based on relevant instrument parameters. Full structural information was then obtained via structural refinements in the order of S (Se) position, the isotropic thermal parameter (B$_{\rm {iso}}$) of Ni, and S (Se) atomic sites. Several other sequences were also tested in the fits, with no meaningful difference detected in the refined crystal structures.

More realistic Se-dopings ($x$) than the nominal values in powder samples were determined through three kinds of refinements. First, Vegard's law between \NS~(\textit{a$_c$}~=~5.6873~\AA)~\cite{ANDRESEN_LP} [close to our single crystal XRD results (not shown)] and \NSe~(\textit{a$_c$}~=~5.9629~\AA)~\cite{ANDRESEN_LP} was utilized to estimate the actual doping, followed by the Rietveld refinements to extract structural parameters. Second, the relative ratios of S and Se occupancies were fitted, assuming the full occupancy in the same $8c$ Wyckoff site. The deviation of refined Se-dopings ($x$) from the values obtained with Vegard's law was only within 0.018 over the entire powder sample, which was much smaller than a typical step of nominal $x$ values ($\delta x$~=~0.05 or 0.1) in our series of powder samples. Last, occupancies of chalcogen ions were refined independently although their fits were least stable; however, nearly same structural parameters (i.e., \dNX, \dXX~and ${a_c}$) were obtained, comparable with those using Vegard's law [as shown in Figs.~\ref{XRD_doping}(b-d)]. The high quality of our power sample made this fit (with free occupancy parameters) more reliable [see very small values of refinement parameters, R$_{\rm p}$, R$_{\rm wp}$ and R$_{\rm exp}$ in~\Cref{table:RT,table:TD}]: in fact, many of refinements parameters are even comparable with those from standard LaB$_{6}$ powder samples (R$_{\rm p}$~$\sim$~14.7; R$_{\rm wp}$~$\sim$~11.9) (not shown). As a result, a more realistic Se doping scheme was adopted based on Vegard's law (the first method). We assumed that full site occupancies in all XRD data with $x$~=~0.24, 0.43, 0.47, 0.49, 0.52, 0.60, and 0.68 (deviated from the nominal doping of $x$ = 0.2, 0.4, 0.45, 0.45, 0.5, 0.6, and 0.7, respectively) [see a full structural information from the refinement in Tables~\ref{table:RT} and~\ref{table:TD}]. This in-depth analysis makes the anomalous trend of structural parameters shown in Figs.~\ref{XRD_doping}(b-d) in the main text more convincing.

All doping-dependent XRD data collected at the 298 and 12~K are presented in Figs.~\ref{app:XRD:RT} and~\ref{app:XRD:LT}, respectively. The most prominent change in the diffraction pattern in terms of Se doping was in the relative intensities of the three strong peaks from 30$^{\circ}$ $\leq$ 2$\theta$ $\leq$ 40$^{\circ}$ surrounded by dashed black lines. This enables more reliable structural refinements.

All temperature-dependent XRD data for three powder samples ($x$~=~0.49, 0.47 and 0.24) are shown in Fig.~\ref{app:XRD:0p49} to Fig.~\ref{app:XRD:0p24}, respectively. In the experiment, data quality was checked at 292~K, followed by the collection of the full diffraction data set as the samples were heated from 12 to 292~K. In the refinement process, ${a_c}$ extracted from the full XRD data collected at 298~K [filled symbols in Fig.~\ref{fig:LP}(d)] was used as a starting value to obtain ${a_c}$ parameters of the remaining temperature data [empty symbols in Fig.~\ref{fig:LP}(d)] in the subsequent~\textit{Le Bail} fit. This was cross-checked by refinements starting with 12~K data for $x$~=~0.49, with consistent results.

As a representative sampling of data quality with corresponding analyses, the entire XRD data set is shown for the $x$~=~0.49 sample over a wide temperature range in Fig.~\ref{app:XRD:0p49}. Figure~\ref{app:XRD:0p49}(a) shows the overall change in the diffraction patterns with temperatures. Figure~\ref{app:XRD:0p49}(b) reveals a clear jump in the chosen (3 1 1) Bragg peak in terms of 2$\theta$ through the transition to the lower temperature. Figure~\ref{app:XRD:0p49}(c) shows reliable refinements, capturing a small peak feature near the transition; notably, the double-peak structure of the whole pattern makes the refinement worse due to the coexistence of the metallic and insulating phase [see Fig.~\ref{fig:XRD_T}(b)]. This was further verified in a failed refinement trial with the (3 1 1) peak data collected to the closer transition temperature (not shown). For completeness, the extracted structural information from all XRD data is summarized in Table~\ref{table:RT} and~\ref{table:TD}.

High-quality crystallites of our powder samples were further corroborated with preliminary synchrotron powder XRD data (not shown). The fitted lattice parameters from the D8 XRD data collected at the home laboratory were used to fit the synchrotron XRD data, to determine the lower and upper bound of the inhomogeneous Se doping for each sample. Via Vegard's law, tiny inhomogeneity concentrations (less than $\Delta x$~=~0.011) were resolved from the same powder samples ($x$~=~0.43, 0.47 and 0.52) previously measured by the D8 XRD system. Thus, the refined Se doping values shown in this paper are representative with a good accuracy.

\begin{figure}[t]
\begin{center}
\includegraphics[width=\linewidth]{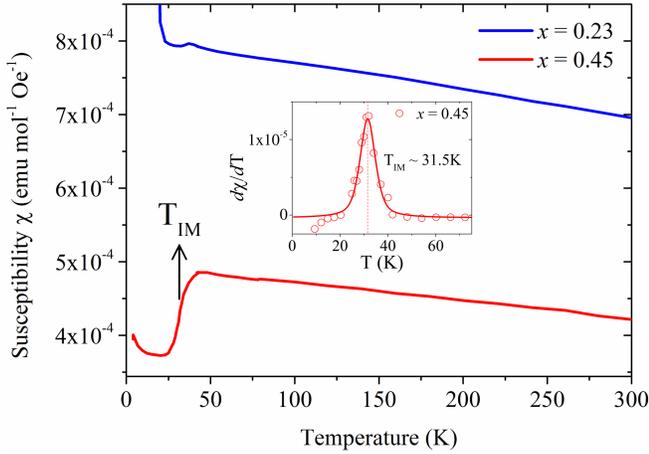}
\end{center} \caption {(color online) Magnetic susceptibility measurements on two single crystals, $x$~=~0.23 and 0.45. \TIM~indicates the onset of the transition derived by the differentiation of susceptibility. In the inset, the same differentiation method is used to obtain the transition temperature with the resistivity data for consistency [see the inset of Fig.~\ref{fig:LP}(c)].}
\label{app:fig:Raman:mpms}
\end{figure}

\begin{figure*}
\begin{center}
\includegraphics[width=\linewidth]{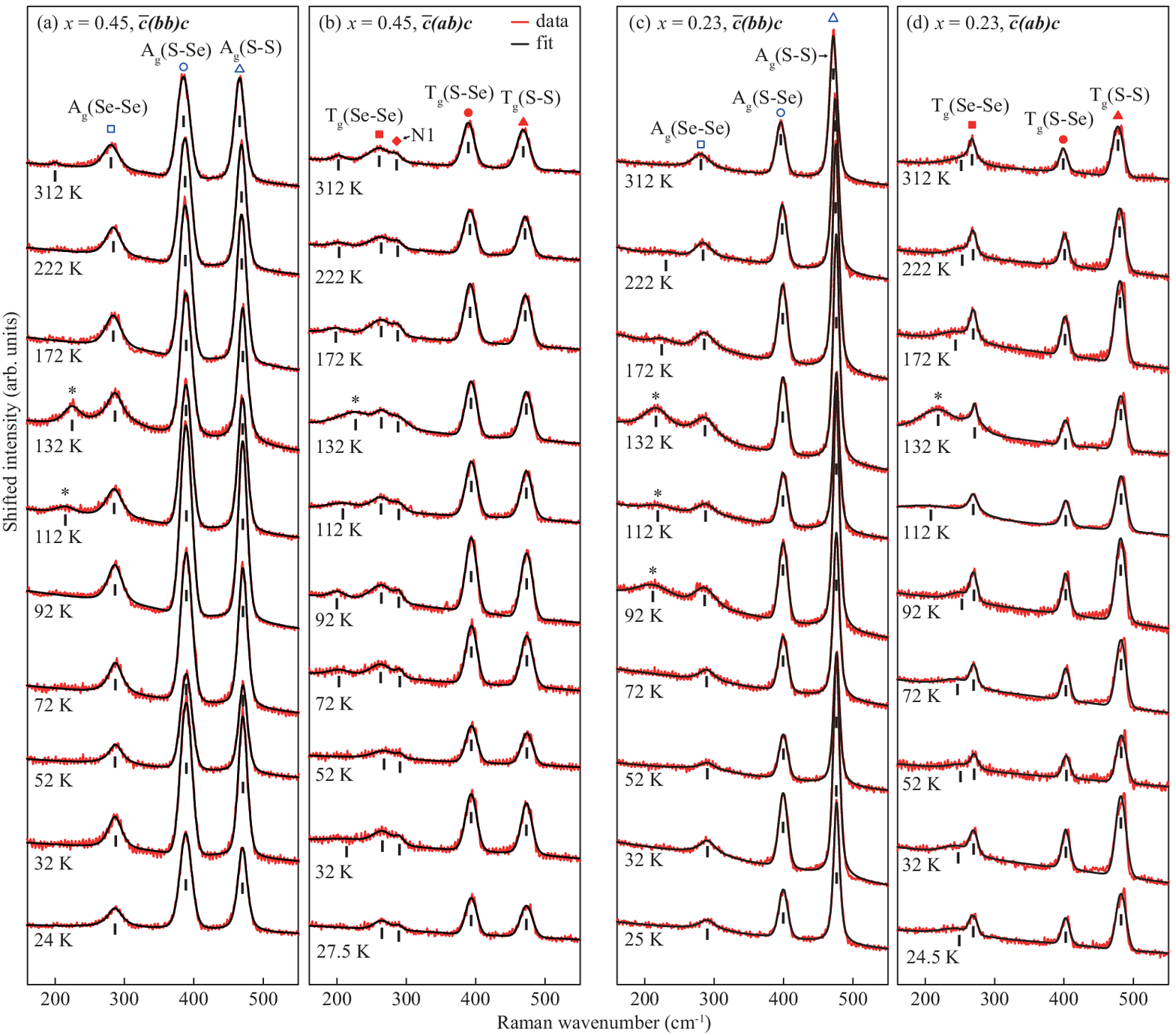}
\end{center} \caption {(color online) Raman spectra for the parallel and perpendicular polarization of $x$~=~0.45 and 0.23. Asterisks indicate ice peaks at about 220\cm. Other symbols for the main peaks in this figure corresponds to those in Figs.~\ref{fig:Raman:TD}(e-g).}
\label{app:Raman:TD}
\end{figure*}

\section{Characterizations of single crystals used for Raman measurements}
\label{app:Raman:char}
Two single crystals ($x$~=~0.23 and 0.45, deviated from nominal 0.2 and 0.45 dopings, respectively) studied by Raman scattering were also characterized by Magnetic Property Measurement System (from Quantum Design), supplemented by four-circle single crystal XRD measurements. Magnetic susceptibility data in Fig.~\ref{app:fig:Raman:mpms} showed a clear signature for the IM transition at about 31.5~K in $x$~=~0.45~whereas this feature was absent in $x$~=~0.23 with two reproduced long-range magnetic transitions~\cite{Matsuura2000}.

More realistic dopings of $x$~=~0.23 and $x$~=~0.45 were obtained from the four-circle single crystal XRD measurements using Vegard’s law with a fitted lattice parameter \textit{a$_c$}, based on two nuclear Bragg peaks of (2 0 0) and (4 0 0). In particular, the Se doping of the $x$~=~0.45 crystal was consistently estimated with a fitted solid black line connecting extracted transition temperatures in Fig.~\ref{fig:LP}(b) obtained from resistivity and high-resolution XRD measurements of powder samples.

\section{Raman Scattering}
\label{app:Raman}
Raman spectra were collected with a Jobin Yvon Typ V 010 LabRAM spectrometer~\cite{Taheri2016} (with the linearly polarized 632.817~nm red light from a He/Ne gas laser) using high-resolution 1800~lines/mm grating monochromator with an instrumental resolution of about 1.8~cm$^{-1}$ (measured with a standard Neon lamp). Lower temperature was attained using a CryoVac crostat with a copper cold finger. Single crystals were mounted with a Leit-Silver paste to ensure the fast and reliable thermal equilibrium between room-temperature and the base temperature ($\sim$~10~K in the cryostat as measured by a Lake Shore temperature controller). The low temperatures were also checked using the already-known Raman spectra from standard materials; i.e., Si~\cite{Hart1970}, GaP~\cite{Ves2001} using a general hardening of phonon peaks with cooled temperatures; SrTiO$_{3}$~\cite{Ouillon:STO:2002} using a new peak that only appears below about 94~K. The Raman beam was focused to a circular spot (a few micrometer in the diameter) on the sample's surface using a Nikon 50x lens. Note that the green 514~nm light from an argon+krypton ion laser (Innova i70c-spectrum) with a JobinYvon T64000 spectrometer was also used, finding no noticeable difference in the Raman spectra at room-temperature.

Importantly, to more accurately determine the wavenumbers of phonon peaks, a secondary Rayleigh filter was equipped to further suppresses an elastic scattering peak. This specialized setup allows the simultaneous measurement of an elastic peak and inelastic peaks. In the collected data, the zero-energy peak was first fitted and shifted to the nominal 0\cm~position in the spectrometer, and the positions of inelastic peaks were calibrated accordingly. The insignificant change in the Raman signal with the secondary Rayleigh filter was also confirmed, as this effectively cuts off signals only close to the elastic line. Typical measurements were performed longer than 30~minutes, and much longer measurements (i.e., 8.5 hours) were carried out to confirm very weak signals for some selected data. To optimize the signal-to-noise signals, the Raman beam was focused laterally with the microscope at each temperature. Nearly the same beam position on the crystal was used for all temperature data. Raman signals from various positions on the crystals did not exhibit significant differences, and this homogeneous chemical composition was further supported by scanning electron microscope and energy-dispersive x-ray spectroscopy analyses (not shown).

Polarized measurements enable Raman peaks to be distinguished by the symmetry argument. The point group T$_{h}$ ($m\bar{3}$) [with a space group \sg~(No.~205)] predicts five Raman-active modes, factorized as A$_g$~+~E$_g$~+~3T$_g$~\cite{Suzuki}. The point group determines the Raman intensity tensor depending on the given directions of light and polarization. For example, the Raman intensity tensor matrices~\cite{Loudon,Suzuki} with X~=~[1 0 0], Y~=~[0 1 0] and Z~=~[0 0 1] are given as
\begin{equation}
I_{001}(A_{g})=
\begin{pmatrix}
${\cal A}$ &         &  \\
         &${\cal A}$ &  \\
         &         &${\cal A}$
\end{pmatrix} \nonumber
\end{equation}

\begin{equation}
I_{001}(E_{g})=
\begin{pmatrix}
${\cal B}$ &         &  \\
         &${\cal B}$ &  \\
         &         &${\cal B}$
\end{pmatrix} \label{eq:Raman}
\end{equation}

\begin{equation}
I_{001}(T_{g})=
\begin{pmatrix}
         &${\cal D}$   &${\cal D}$ \\
${\cal D}$  &          & ${\cal D}$ \\
${\cal D}$  &${\cal D}$  &
\end{pmatrix}, \nonumber
\end{equation}
where A, B and D are Raman intensity components and a subscript indicates the direction of light, where [0 1 0] [as given in the literature~\cite{Suzuki}] was changed to an equivalent [0 0 1] direction in the cubic symmetry, to make it more directly connecting with our experimental notations.

With the backscattering geometry, we employ a Porto's notation~\cite{Damen1966} to describe the direction and polarization of incident and scattered light in the form of $k_{i}(E_{i} E_{s})k_{s}$, where $k_{i}$ ($k_{s}$) is the direction of incident (scattered) light and $E_{i}$ ($E_{s}$) is the polarization of incident (scattered) light. Two polarizations [parallel $\bar{\bm c}(\bm b \bm b){\bm c}$~and perpendicular $\bar{\bm c}(\bm a \bm b){\bm c}$~polarizations] were used, where ${\bm a}$, ${\bm b}$ and ${\bm c}$ indicate the cubic crystallographic directions. This notation was used instead of X, Y and Z to avoid any confusion with other notations in this paper, and a bar indicates the negative direction.

For completeness, the full Raman data set is shown in Fig.~\ref{app:Raman:TD}. In addition to stretching phonon modes [see Figs.~\ref{fig:Raman:TD}(e-g)], a few other weak peaks were observed; however, such peaks could not be readily identified. For instance, the tiny peaks around 200~meV are too close to a stronger ice signal nearby. Additionally, weak shoulder peaks appeared at a lower wavenumber of A$_{g}$(Se - Se) peaks; these features may be related to the rotation modes (such as E$_{g}$) of S - S bonds~\cite{Heras} energetically close to the stretching modes of Se - Se.

Raman data were fit using a convoluted pseudo-Voigt function of Lorentzian and Gaussian profiles~\cite{Marini2014} to quantitatively describe phonon peaks by extracting more accurate peak positions. A more conventional Raman fitting method~\cite{Marini:fit} was also applied, to take into account the electronic response; however, this was not successful, most likely due to much weaker broad continuous signals below the main phonon peaks, especially towards the lower wavenumber in the polarized Raman data than the unpolarized counterpart. Thus, we focused more on extracting phonon wavenumbers with two representative crystals ($x$~=~0.23 and 0.45), to locate any phonon anomaly related to the transition with temperature and doping variations. Fitting with the convoluted pseudo-Voigt function works much better with an empirical linear background combined with a constant background [visualized as solid black lines in Figs.~\ref{fig:Raman:TD}(a-d) and Fig.~\ref{app:Raman:TD}].

\begin{figure}
\begin{center}
\includegraphics[width=\linewidth]{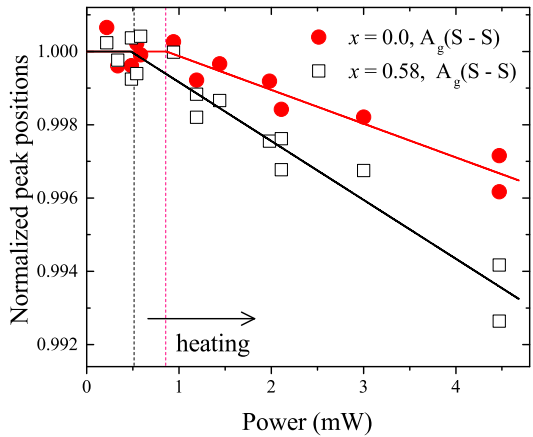}
\end{center}\caption {(color online) (a) Normalized averaged peak positions of A$_{g}$ mode (S - S) from Raman measurements on two single crystals ($x$~=~0, 0.58). The two crystals exhibited different softening rates with laser power guided by the fitted solid curves. The threshold power, at which the phonon softening is initiated, is indicated by vertically dotted lines at 0.5~mW (a lower bound) and 0.85~mW (an upper bound) for $x$~=~0.58 and 0, respectively.}
\label{app:fig:Raman:BH}
\end{figure}

\section{Beam-Heating in Raman Scattering}
\label{app:Raman:BH}
Independent Raman measurements were conducted to estimate the beam-heating effect by tracing the shift of phonon peaks with laser power at room-temperature. With two representative single crystals [$x$~=~0 and 0.58, which were also characterized by susceptibility and four-circle single crystal XRD measurements (not shown)], we chose the A$_{g}$(S - S) phonon in a parallel polarization $\bar{\bm c}(\bm b \bm b){\bm c}$~due to a stronger intensity and more symmetric peak shape than the others. Their peak positions were normalized by the reference phonon wavenumber measured with the lowest Raman laser power, in which beam heating is clearly not present.

Figure~\ref{app:fig:Raman:BH} illustrates the relative peak positions (symbols) and their fits (solid lines), showing a nearly flat behavior from zero power up to 0.8~mW for $x$~=~0.0 (0.5~mW for $x$~=~0.58 as a minimum level of power without heating), followed by the onset of a gradual softening of phonons owing to beam heating. The data revealed that the beam-heating effect seems to occur similarly from 0.8~mW for both samples, indicating a general trend of it over the wide range of Se doping up to $x$~=~0.58. However, we should mention that it is rather difficult to determine the precise laser power at which beam heating clearly begins; thus, the lower-bound value of the laser power with the $x$~=~0.58 data is shown in Fig.~\ref{app:fig:Raman:BH} as a reference. The complete list of peak positions used for this analysis is given in Table~\ref{table:Raman:BH}.

In the Raman data shown in Fig.~\ref{fig:Raman:TD} in the main text, 2.11~mW was used (except for the lowest temperature data collected at 1.19~mW), which allows a clear detection of even weak Raman peaks, such as A$_{g}$(Se - Se) and T$_{g}$(Se - Se) modes between 270 and 290~cm$^{-1}$. This is especially helpful in tracing the evolution of Se - Se phonons with temperature and testing the role of the Se - Se dimer in the transition~\cite{XAS:HJN}. We estimated an averaged beam-heating of $\sim$~20~K ($\sim$~13~K for 1.19~mW setting), assuming a linear evolution of phonon peaks from room-temperature down to 202~K. This is a reasonable assumption based on a similarly observed linear trend in phonon shifts with temperature [Figs.~\ref{fig:Raman:TD}(e,g)]. The temperatures in Fig.~\ref{fig:Raman:TD} and Fig.~\ref{app:Raman:TD} were calibrated accordingly.

\begin{table*}
\caption[] {Fitted positions of phonon peaks with fitting errors from $\bar{\bm c}(\bm b \bm b){\bm c}$ and $\bar{\bm c}(\bm a \bm b){\bm c}$ polarization data of $x$~=~0.23 (upper panel) and $x$~=~0.45 (lower panel), respectively. The lowest temperature measured varied slightly: 25~K in $\bar{\bm c}(\bm b \bm b){\bm c}$, 24.5~K in $\bar{\bm c}(\bm a \bm b){\bm c}$ for $x$~=~0.23 and 24~K in $\bar{\bm c}(\bm b \bm b){\bm c}$, 27.5~K in $\bar{\bm c}(\bm a \bm b){\bm c}$ for $x$~=~0.45.}
\label{table:Raman}
\par
\vspace{-0.4cm}
\begin{center}%
\begin{tabular}
[c]{c c c c c c c c c c c c c}\hline\hline T (K) &~A$_{g}$(Se - Se) &~A$_{g}$(S - Se) &~A$_{g}$(S - S) &~T$_{g}$(Se - Se) &~T$_{g}$(S - Se) &~T$_{g}$(S - S) \\
\hline
24 &~289.86(39) &~399.29(4) &~476.10(1) &~268.69(2) &~402.04(1) &~481.91(1) \\
32 &~290.43(9) &~399.39(1) &~476.08(0) &~269.42(3) &~402.31(2) &~481.62(1) \\
52 &~290.04(12) &~399.61(1) &~476.30(1) &~270.32(15) &~402.73(8) &~482.33(4) \\
72 &~289.34(34) &~399.44(4) &~476.01(1) &~269.52(3) &~402.25(2) &~481.76(1) \\
92 &~286.31(10) &~399.37(1) &~475.94(0) &~269.40(4) &~402.14(2) &~481.34(1) \\
112 &~287.64(12) &~399.42(1) &~476.02(0) &~268.74(3) &~402.22(2) &~481.84(1) \\
132 &~286.88(12) &~399.24(1) &~475.89(0) &~270.81(2) &~401.96(2) &~481.54(1) \\
172 &~285.97(13) &~398.76(1) &~475.15(0) &~268.93(2) &~401.11(1) &~480.42(1) \\
222 &~283.90(9) &~398.19(1) &~474.44(0) &~268.56(2) &~400.61(1) &~479.76(1) \\
312 &~280.93(7) &~395.87(1) &~471.74(0) &~267.27(1) &~398.59(1) &~477.33(1) \\
\hline\hline
\end{tabular}
\end{center}
\par
\vspace{-0.6cm}
\begin{center}
\begin{tabular}
[c]{c c c c c c c c c c c c c c c}\hline\hline T (K) &~A$_{g}$(Se - Se) &~A$_{g}$(S - Se) &~A$_{g}$(S - S) &~T$_{g}$(Se - Se) &~N1 &~T$_{g}$(S - Se) &~T$_{g}$(S - S) \\
\hline
24 &~286.25(4) &~388.02(1) &~469.52(1) &~264.62(7) &~288.97(9) &~393.53(1) &~473.24(1) \\
32 &~286.38(5) &~388.59(1) &~469.91(1) &~265.31(23) &~290.04(20) &~393.39(2) &~473.35(2) \\
52 &~287.28(34) &~388.77(4) &~470.13(5) &~267.94(37) &~290.72(32) &~393.99(3) &~473.95(4) \\
72 &~286.95(5) &~388.71(1) &~469.97(1) &~262.94(15) &~290.60(13) &~393.87(2) &~473.71(2) \\
92 &~286.21(4) &~388.41(1) &~469.73(1) &~263.99(11) &~289.19(8) &~393.47(1) &~473.45(1) \\
112 &~285.22(5) &~388.26(1) &~469.45(1) &~263.40(14) &~289.09(15) &~393.67(1) &~473.25(1) \\
132 &~286.29(4) &~388.42(1) &~469.63(1) &~264.06(13) &~287.97(22) &~393.25(1) &~473.05(1) \\
172 &~283.83(3) &~387.20(1) &~468.18(1) &~263.58(12) &~287.60(8) &~391.91(1) &~471.65(1) \\
222 &~283.99(4) &~386.97(1) &~467.91(1) &~263.52(13) &~287.80(10) &~391.62(1) &~471.37(1) \\
312 &~280.46(3) &~384.68(1) &~465.32(1) &~260.87(11) &~286.11(11) &~389.37(1) &~468.54(1) \\
\hline\hline
\end{tabular}
\end{center}
\end{table*}

\begin{table*}
\caption[] {Structural parameters extracted from \NSS~powder x-ray data at 298~K and 12~K for a wide range of representative dopings (\sg~space group, No. 205, $Z=4$). B$_{\rm iso}$ is the isotropic thermal displacement. Note that 292~K was used for $x$~=~0.24, ~0.47 and ~0.49.}
\label{table:RT}
\par
\vspace{-0.4cm}
\begin{center}
\centering
\begin{tabular}
[c]{c c c c c c c c}\hline\hline  T~=~298~K & $x$~=~0.24 & $x$~=~0.43 & $x$~=~0.47 & $x$~=~0.49 & $x$~=~0.52 & $x$~=~0.60 & $x$~=~0.68 \\
\hline
R$_{\rm p}$                          ~ &23.3          ~ &19.7              ~ & 23.8~             &18.9~             &17.7~            &17.1~           &18.9\\
R$_{\rm wp}$                       ~ &11.2          ~ &11.2               ~ & 12~               &10~                &9.72~            &9.35~           &10.5\\
R$_{\rm exp}$                      ~ &9.42          ~ &8.41               ~ &9.68  ~            &8.31~             &7.55~            &7.97~           &7.83\\
a (\AA)                   ~ & 5.71947(2) ~ & 5.74607(2)    ~  &5.75279(2) ~  &5.75434(2)~   &5.75829(2)~  &5.76976(2)~  &5.78065(2)\\
x (S, Se)                ~ & 0.39376(6) ~ & 0.39193(5)    ~  &0.39207(6) ~   &0.39153(4)~   &0.39108(4)~  &0.39071(5)~  &0.38961(5)\\
B$_{\rm iso}$~(Ni)  ~ &0.742(18) ~ &0.419(17)         ~ &0.382(20) ~     &0.593(17)~     &0.832(15)~    &0.726(16)~   &0.936(19)\\
B$_{\rm iso}$~(X)   ~ &0.704(16) ~ &0.378(14)         ~ &0.582(17) ~     &0.564(13)~     &0.788(12)~    &0.440(12)~   &0.349(13)\\
\dNX                      ~ &2.4105(4) ~ &2.4172(3)         ~ &2.4204(4) ~     &2.4197(3)~     &2.4203(3)~    &2.4243(3)~   &2.4263(3)\\
\dXX                      ~ &2.1049(5)  ~ &2.1511(5)        ~ &2.1509(5)~      &2.1622(5)~     &2.1727(5)~    &2.1844(5)~   &2.2105(5)\\
\hline\hline
\end{tabular}
\end{center}
\par
\vspace{-0.6cm}
\begin{center}
\begin{tabular}
[c]{c c c c c c c c} \hline\hline   T~=~12~K & $x$~=~0.24 & $x$~=~0.43 & $x$~=~0.47 & $x$~=~0.49 & $x$~=~0.52 & $x$~=~0.60 & $x$~=~0.68 \\
\hline
R$_{\rm p}$ ~                        &22.4  ~           &19.3~           & 24.5~ &17.8~ &16.8~ &16.4~ &18.1\\
R$_{\rm wp}$ ~                      &11  ~              &11~             & 13~ &9.65~ &9.39~ &9.2~ &10.6\\
R$_{\rm exp}$ ~                     &9.13  ~           &8.23~          & 9.41~  &8.06~ &7.31~ &7.79~ &7.59\\
a (\AA) ~                  & 5.70709(2)  ~ & 5.73287(2)~ & 5.72926(2)~ & 5.73077(1)~ & 5.73446(1)~ & 5.74857(2)~ & 5.76165(2)\\
x (S, Se) ~               & 0.39362(6)  ~ & 0.39201(5)~ & 0.39165(7)~ & 0.39120(4)~ & 0.39122(4)~ & 0.39017(4)~ & 0.38917(5)\\
B$_{\rm iso}$~(Ni) ~ &0.458(17)~     &0.110(16)~ &0.132(20) ~&0.283(15)~ &0.544(14)~ &0.334(15)~ &0.577(18)\\
B$_{\rm iso}$~(X) ~ &0.488(15) ~    &0.173(13)~ & 0.344(17)~&0.310(12)~ &0.503(11)~ &0.162(11)~ &0.094(12)\\
\dNX~                     &2.4049(4)~     &2.4119(3)~ &2.4095(5)~  &2.4091(3) ~ &2.4107(3)~ &2.4141(3)~ &2.4173(3)\\
\dXX~                     &2.1031(5)~     &2.1446(5)~ &2.1504(6)~  &2.1599(5) ~ &2.1609(4)~ &2.1871(4)~ &2.2120(5)\\
\hline\hline
\end{tabular}
\end{center}
\end{table*}

\clearpage

\begin{table*}
\caption[] {Extracted structural parameters from temperature-dependent XRD data for $x$~=~0.47, ~0.49 and ~0.24}
\label{table:TD}
\par
\vspace{-0.4cm}
\begin{center}
\begin{tabular}
[c]{c c c c c c c}\hline\hline $x$~=~0.47 & 12~K & 32~K & 62~K & 92~K & 152~K & 292~K \\
\hline
R$_{\rm p}$ &  24.5  &      24.7 &        22.1 &        21.9  &       22.6 &        23.8 \\
R$_{\rm wp}$ & 13 &13.8 &11.1 &11.1 &11.5 &  12\\
R$_{\rm exp}$ & 9.41 & 9.4 &9.36 &9.38 &9.44 &9.68\\
a (\AA) & 5.72926(2) & 5.72955(2) & 5.74190(2) & 5.74285(2) & 5.74541(2) & 5.75279(2)\\
x (S, Se) & 0.39165(7) &0.39165(6) &0.39171(6) &0.39179(6) &0.39179(6) &0.39207(6) \\
B$_{\rm iso}$~(Ni) & 0.132(20) &0.142(22) &0.153(18) &0.123(17) & 0.211(18) &0.382(20) \\
B$_{\rm iso}$~(X) & 0.344(17) &0.336(18) &0.338(15) &0.321(15) &0.385(16) &0.582(17) \\
\dNX~ & 2.4095(5) &2.4096(5) &2.4149(4) &2.4155(4) &2.4166(4) &2.4204(4) \\
\dXX~ & 2.1504(6) &2.1505(6) &2.1539(5) &2.1527(5) &2.1537(5) &2.1509(5) \\
\hline\hline
\end{tabular}
\end{center}
\par
\vspace{-0.6cm}
\begin{center}
\begin{tabular}
[c]{c c c c c c c c c c}\hline\hline $x$~=~0.49  & 12~K & 32~K & 52~K & 72~K & 92~K & 112~K & 152~K & 202~K & 292~K \\
\hline
R$_{\rm p}$ &17.8 & 18 &	18	 &20.1 &	19.1 ~ &18.8~ &	18.4~ &	18.8~ &	18.9 \\
R$_{\rm wp}$ &9.65 &9.8 &	9.96 &	11.6 &	10.4 ~ &10.3	~ &10.1	~ &10.2~ & 10\\
R$_{\rm exp}$ &8.06 & 8.08 &	8.08 &	8.12 &	8.11 ~ &8.13~ &	8.17~ &	8.2~ &	8.31\\
a (\AA) & 5.73077(1) & 5.73102(1) & 5.73178(1) & 5.74333(2) & 5.74454(2) & 5.74538(2) & 5.74708(2) & 5.74968(2) & 5.75434(2) \\
x (S, Se) &0.39120(5)&	0.39115(5)&	0.39119(5)&	0.39125(5)&	0.39133(5) ~ & 0.39131(5)~ & 	0.39140(5)~ & 	0.39138(5)~ & 0.39153(5) \\
B$_{\rm iso}$~(Ni)&0.283(15) ~&0.298(15) ~& 	0.287(15) ~& 	0.345(18) & 	0.353(16) ~ &    0.371(16)~ &	0.405(16)~ &	0.463(16)	~ &0.593(16)\\
B$_{\rm iso}$~(X)&0.310(12) ~& 0.314(12) ~& 	0.300(12) ~& 	0.349(14) & 	0.348(13) ~ &0.378(13)~ &	0.396(13)~ &	0.458(13)	~ &0.564(13)\\
\dNX~ & 2.4091(3)&2.4090(3)&	2.4095(3)&	2.4145(4)&	2.4151(3) ~ & 2.4155(3)~ & 	2.4164(3)~ & 	2.4174(3)~ & 	2.4197(3) \\
\dXX~ & 2.1599(5)&2.1610(5)&	2.1605(5)&	2.1636(5)&	2.1625(5)~ & 2.1632(5)~ & 	2.1621(5)~ & 	2.1634(5)~ & 	2.1622(5) \\
\hline\hline
\end{tabular}
\end{center}
\par
\vspace{-0.6cm}
\begin{center}
\begin{tabular}
[c]{c c c c c c c c c c}\hline\hline $x$~=~0.24  & 12~K & 32~K & 52~K & 72~K & 92~K & 112~K & 152~K & 202~K & 292~K \\
\hline
R$_{\rm p}$~&22.4~&22.4~&22.8~&22.4~&22.8~&22.7~&22.8~&22.8~&23.3 \\
R$_{\rm wp}$~&11~&11~&11.3~&11.1~&11.1~&11.1~&11.1~&11.1~&11.2 \\
R$_{\rm exp}$~&9.13~&9.18~&9.15~&9.16~&9.18~&9.19~&9.24~&9.32~&9.42 \\
a (\AA) ~&5.70709(2)~&5.70778(2)~&5.70828(2)~&5.70861(2)~&5.70913(2)~&5.70981(2)~&5.71160(2)~&5.71419(2)~&5.71947(2) \\
x (S, Se) ~&0.39362(6)~&0.39357(6)~&0.39357(6)~&0.39364(6)~&0.39362(6)~&0.39373(6)~&0.39373(6)~&0.39371(6)~&0.39376(6) \\
B$_{\rm iso}$~(Ni)~&0.458(17)~&0.457(16)~&0.484(17)~&0.469(17)~&0.503(17)~&0.523(17)~&0.569(17)~&0.627(17)~&0.742(18) \\
B$_{\rm iso}$~(X) ~&0.488(15)~&0.492(15)~&0.477(15)~&0.503(15)~&0.517(15)~&0.528(15)~&0.570(15)~&0.603(15)~&0.704(16) \\
\dNX~~&2.4049(4)~&2.4051(4)~&2.4053(4)~&2.4056(4)~&2.4058(4)~&2.4063(4)~&2.4071(4)~&2.4081(4)~&2.4105(4) \\
\dXX~~&2.1031(5)~&2.1044(5)~&2.1046(5)~&2.1033(5)~&2.1039(5)~&2.1020(5)~&2.1026(5)~&2.1040(5)~&2.1049(5) \\
\hline\hline
\end{tabular}
\end{center}
\end{table*}

\begin{table*}
\caption[] {Fitted positions of the A$_{g}$ (S - S) phonon with increasing Raman power on two single crystals of $x$~=~0, 0.58. Note that all fitting error bars are smaller than 0.04 for $x$~=~0 and 0.12 for $x$~=~0.58.
}
\label{table:Raman:BH}
\par
\vspace{-0.4cm}
\begin{center}
\begin{tabular}
[c]{c c c }\hline\hline Power (mW) &~$x$~=~0 (cm$^{-1}$)  &~$x$~=~0.58 (cm$^{-1}$)\\
\hline
0.216~&477.80~&461.98 \\
0.337~&477.30~&461.76 \\
0.484~&477.30~&461.52, 462.04 \\
0.543~&477.59~&461.59 \\
0.585~&477.44~&462.06 \\
0.94~&477.61~&461.86 \\
1.19~&477.11~&461.04, 461.33 \\
1.44~&477.32~&461.25 \\
1.98~&477.10~&460.74 \\
2.11~&476.73~&460.38, 460.77 \\
3~&476.63~&460.37 \\
4.47~&475.66, 476.13~&458.47, 459.18 \\
\hline\hline
\end{tabular}
\end{center}
\end{table*}

\end{document}